\documentclass[sigconf]{acmart}

\usepackage{booktabs} 
\usepackage{balance}
\usepackage{multirow}
\usepackage{adjustbox}
\usepackage{xcolor}
\usepackage{hyperref}
\usepackage{mfirstuc}
\usepackage{titlecaps}
\usepackage{stringstrings}
\usepackage{booktabs}
\usepackage{tabularx}
\usepackage{tabularx,booktabs} 
\usepackage{float}

\copyrightyear{2018}
\acmYear{2018}
\setcopyright{acmcopyright}
\acmConference[SoHeal'18]{SoHeal'18:IEEE/ACM 1st International Workshop on Software Health }{May 27-June 3, 2018}{Gothenburg, Sweden}
\acmBooktitle{SoHeal'18: SoHeal'18:IEEE/ACM 1st International Workshop on Software Health , May 27-June 3, 2018, Gothenburg, Sweden}
\acmPrice{15.00}
\acmDOI{10.1145/3194124.3194130}
\acmISBN{978-1-4503-5730-2/18/05}

\newif\ifcommenttrue
\newcommand{\ignore}[1]{  }

\newcommand{\matteo}[1]{\textcolor{blue}{{\it [Matteo says: #1]}}}

\begin{document}
\title{CloudHealth: A Model-Driven Approach to Watch the Health of Cloud Services}

\author{Anas Shatnawi}
\affiliation{%
  \institution{University of Milan-Bicocca}
  \streetaddress{Viale Sarca 336}
  \city{Milan} 
  \country{Italy}
  \postcode{20126}
}
\email{anas.shatnawi@unimib.it}

\author{Matteo Orr\`u}
\affiliation{%
  \institution{University of Milan-Bicocca}
  \streetaddress{Viale Sarca 336}
  \city{Milan} 
  \country{Italy}
  \postcode{20126}
}
\email{matteo.orru@unimib.it}

\author{Marco Mobilio}
\affiliation{%
  \institution{University of Milan-Bicocca}
  \streetaddress{Viale Sarca 336}
  \city{Milan} 
  \country{Italy}
  \postcode{20126}
}
\email{marco.mobilio@unimib.it}

\author{Oliviero Riganelli}
\affiliation{%
  \institution{University of Milan-Bicocca}
  \streetaddress{Viale Sarca 336}
  \city{Milan} 
  \country{Italy}
  \postcode{20126}
}
\email{oliviero.riganelli@unimib.it}

\author{Leonardo Mariani}
\affiliation{%
  \institution{University of Milan-Bicocca}
  \streetaddress{Viale Sarca 336}
  \city{Milan} 
  \country{Italy}
  \postcode{20126}
}
\email{leonardo.mariani@unimib.it}

\renewcommand{\shortauthors}{A. Shatnawi, M. Orr\`u, M. Mobilio, O. Riganelli, L. Mariani}

\begin{abstract}
 Cloud systems are complex and large systems where services provided by different operators must coexist and eventually cooperate. In such a complex environment, controlling the health of both the whole environment and the individual services is extremely important to timely and effectively react to misbehaviours, unexpected events, and failures.
Although there are solutions to monitor cloud systems at different granularity levels, how to relate the many KPIs that can be collected about the health of the system and how health information can be properly reported to operators are open questions. 

This paper reports the early results we achieved in the challenge of monitoring the health of cloud systems. In particular we present CloudHealth, a model-based health monitoring approach that can be used by operators to watch specific quality attributes. 
The CloudHealth Monitoring Model 
describes how to operationalize high level monitoring goals by dividing them into subgoals, deriving metrics for the subgoals, and using probes to collect the metrics. We use the CloudHealth Monitoring Model to control the probes that must be deployed on the target system, the KPIs that are dynamically collected, and the visualization of the data in dashboards.
\end{abstract}


\maketitle
\section{Introduction}
\label{sec:introduction}
During the last decade, the cloud computing paradigm had a pervasive diffusion, and this is witnessed by its huge impact on the information and communication technology landscape, with a revenue for the total market of public services that is estimated to be of 411.4 USD Billions worldwide by the end of 2020, with an increment of 105.6 USD Billions in the next two years \cite{gartner:17}.

By allowing users (both companies and individuals) to have access to computational resources in flexible, scalable and almost ubiquitous way the cloud-based  systems actually contributed to reshape the way IT companies of almost any industries run their business. As a matter of fact, many economic subjects are increasingly turning their old fashion distributed infrastructure in a new cloud-based one. 



Recent technological trends seem to be addressed at overcoming the traditional IaaS paradigm \cite{bhardwaj2010cloud} toward a more mature PaaS framework \cite{ferrer2016multi}, aiming at further enhancing cloud versatility and scalability. In particular, the cloud is evolving toward a platform for the \emph{universal connectivity} of a variety of \emph{actors}, no matter if human or not (e.g., robots, devices, sensors, etc.), crossing the border that separates different realms (e.g.,. mobile, IoT, Telco, etc.) \cite{Alleman:Rappoport:Banerjee:2010}.  
%
In this scenario where high adaptability and full integration of a large number of different services and actors in the same cloud infrastructure is demanded, requirements such as \emph{configurability} and \emph{programmability} are prominent. However, the high flexibility of the cloud platform must not compromise the \emph{general health} of the systems, which have to satisfy strict reliability and availability requirements.

Addressing this trade-off between flexibility and health is particularly complex in cloud systems because it requires controlling and observing the behaviour of a plethora of sub-systems of different kinds and origin, including both virtual and physical equipments, and dealing with a large variety of subjects (e.g., business actors such as cellular, network and cloud operators, network equipment vendors, application developers) each one with different goals and objectives. 
%
%
%
%
%
While there are a number of engineered solutions to control resource and service allocations in the cloud ~\cite{Szabo:Kind:Westphal:Woesner:Jocha:Csaszar:2015}, 
continuously checking the health of a cloud system is still an open issue. There are already solutions to collect Key Performance Indicators (KPIs) of any type from any resource and service, and there is also the technology to dynamically deploy the probes that can collect these KPIs. However, there is little work about linking these KPIs together into a \emph{general framework} that can \emph{visualize the health status} of both the \emph{system} and its \emph{individual components} based on \emph{operator-specific} objectives.

Capturing the health status of a cloud system poses several challenges: (i) relating the concept of healthiness to actual KPIs in a way that is satisfactory for every actor involved in cloud systems, (ii) reporting the information about health and corresponding KPIs appropriately to cloud users, and (iii) being able to dynamically reconfigure the concept of health of the system, the KPIs that must be collected, and the deployed probes, based on dynamically emerging needs.     

Note that observing the health of the system is not only important for cloud operators, who can manually intervene as soon as problems are detected, but is also necessary to many approaches for self-healing and automatic program repair~
\cite{Ghosh:Sharman:Raghav:Upadhyaya:2007,Psaier:Dustar:2011,Schneider:Barker:Dobson:2015,Al-oqily:Bani-Mohammad:Subaih:Alshaer:2012, Gazzola:Micucci:Mariani:2017}. 

%




In this paper we present the initial results that we obtained in the design of \emph{CloudHealth}, a model-driven approach for monitoring the health of cloud systems. CloudHealth uses a model inspired by the ISO 25011 standard~\cite{ISO25011} to encode the concept of health of a cloud system and to relate this concept to metrics that can be measured on the actual system. This model is used to control the probes that must be deployed on the target system, the KPIs that are dynamically collected, and the visualization of the data in dashboards. 

Interestingly, this model also represents the basis for the definition of a common language that can improve the communication between the many parties involved in the operation of a cloud system. Although we provide an initial definition of this model, the model can be changed at any time to precisely capture the goals and best-practices of specific sub-communities and sets of operators. For instance, what the health of a system is may depend on the stakeholders and the business goals of the operators.  

The paper is organized as follows. Section~\ref{sec:overview} overviews the CloudHealth approach. Section~\ref{sec-modeldriven-monitoring-goals} presents the CloudHealth Monitoring Model. Section~\ref{sec:process} describes the approach in details. Section~\ref{sec:useCases} illustrates some use cases. Section~\ref{related} discusses related work. Section~\ref{sec.conclusion} provides final remarks.

\begin{figure*}[h]
	\begin{center}
		\includegraphics[width=\textwidth]{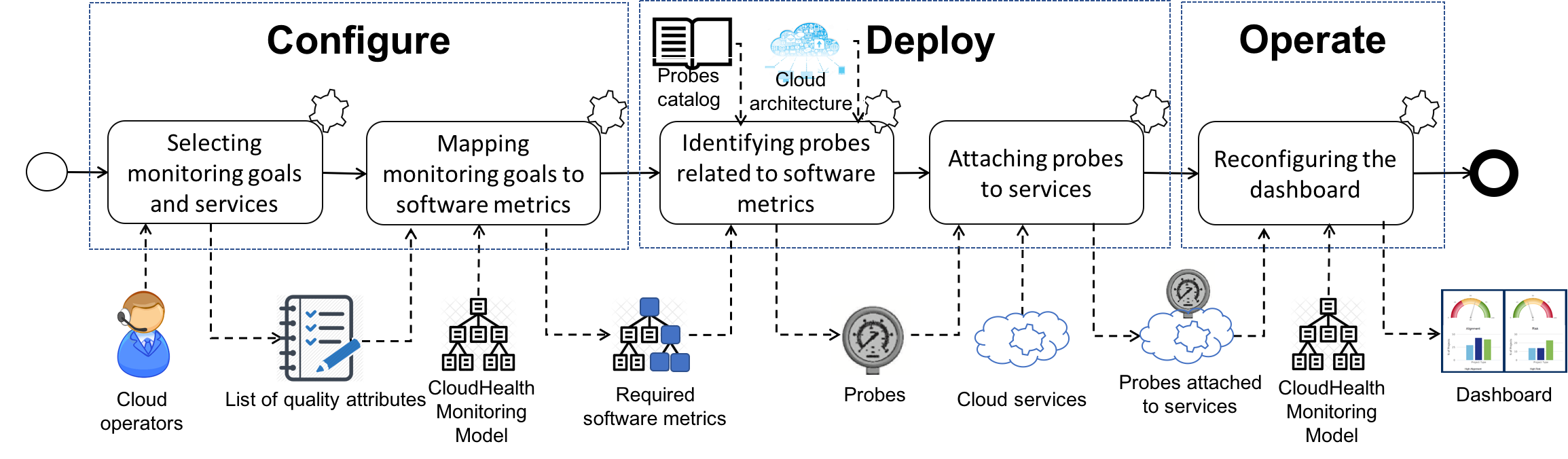}
		\caption{The CloudHealth monitoring process.}
		\label{fig:process}
        \end{center}
\end{figure*}

\newcommand{\CloudHealth}{\begin{verbatim}CloudHealth\end{verbatim}}

\section{\protect\titlecap{CloudHealth}} 
\label{sec:overview}

CloudHealth aims at providing cloud users with a solution to configure, deploy and operate the monitoring infrastructure. CloudHealth provides \emph{data probes}, \emph{data collection} and \emph{data analysis} as a service that collectively result into a built-to-order monitoring environment.  As shown in Figure~\ref{fig:process}, the CloudHealth process is structured in three main stages. 



In the \textit{Configure} stage, the cloud operator selects the \emph{monitoring goals}, that are, the quality attributes that must be computed and observed during operation, and the relevant cloud \emph{services}, that are, the set of services that must contribute to the quality attributes. For example, the cloud operator may decide to monitor the reliability and the efficiency of a subset of the services available in the target platform.

The choice made by the operator is automatically \emph{mapped} into a set of metrics that must be collected from the selected services. The mapping is achieved by using the \emph{CloudHealth Monitoring Model}, which specifies how high level quality attributes can be measured from metrics monitored on individual services. 
 For example, if we consider \textit{reliability} as a quality attribute, this is characterized by three sub-attributes such as continuity, recoverability and availability. The latter, for example, can be analyzed by monitoring metrics such as the \textit{duration of failures and the number of failure occurrences}. 



In the \textit{Deploy} stage, the list of metrics that must be collected are mapped to a set of probes to be deployed on the target system based on the knowledge of the \emph{architecture} of the system and a \emph{probes catalog} with all the probes that can be deployed. 
For example, if the services to be monitored are executed on virtual machines and the time required to recover from failures
must be measured, a probe that performs heartbeat pings (such as pings via ICMP or TCP) can be used. 
The identified probes can then be deployed and \emph{attached} to the monitored services. 

In the \textit{Operate} stage, a dashboard is automatically \emph{reconfigured} according to the goals selected by the operator that can be used to watch the health of both the system and its services. Operators can keep track of the system behaviour by looking at a series of different kind of visualizations, at different level of granularity (goal, sub-goals, cloud properties). For instance, the operator who is interested in monitoring system performance can visualize not only if the system performance is under control, but also the different component of Performance, all the way down to the leaves, where we find Throughput, Latency and Response Time.
%
%
The dashboard is the starting point for problem diagnosis. For instance, if a monitoring goal takes an unhealthy state, the related KPIs, which represent software metrics measured on specific resources, can be shown for further analysis.

Note that after the initial selection of the quality attributes and the services to be observed, the rest of the process can be performed automatically. In this way, CloudHealth can let monitoring tools play an important role in supporting the management of cloud systems to align them with real business needs by providing a user-centered monitoring view that can flexibly and easily change over time.


\section{CloudHealth Monitoring Model}
\label{sec-modeldriven-monitoring-goals}

The aim of the CloudHealth Monitoring Model (CHMM for short) is to represent the relations between high level monitoring goals and the actual software metrics that can be collected from cloud services. Cloud operators can exploit high-level monitoring goals to quickly and easily determine the aspects that must be taken under control using the dashboard. The mapping to low level metrics facilitates and automates the deployment of the probes. The model also specifies how the high-level monitoring goals can be computed from the individual metrics collected at runtime to enable the inspection of the data at different levels of abstractions.  

In order to define the CHMM, we have been inspired by the ISO/IEC 25010:2011 \cite{ISO25010} and ISO/IEC TS 25011:2017 \cite{ISO25011} standards. We selected these standards because they provide \textit{de facto} specifications of quality models to evaluate the quality of general IT services and they can be easily adapted to cloud services. In practice, we identified the quality attributes of the standards that can be used as monitoring goals by cloud operators. Then, we followed the standards to refine these goals into subgoals that can be mapped into measurable cloud service proprieties. We finally identified the metrics that can be used to measure these properties.

\subsubsection*{Defining Monitoring Goals}
Monitoring goals are high level monitoring objectives that can be selected by cloud operators (e.g., Performance). 
We identify the monitoring goals by studying three quality models offered by the 
ISO standards: (i) the \textit{IT service quality}, (ii) the \textit{product quality} and (iii) the \textit{quality in use} models. 
Monitoring goals are identified based on quality attributes that could be adapted for cloud services. We examined 
 each quality characteristic based on the definition offered by the ISO standards 
and selected seven monitoring goals based on seven quality attributes which are \textit{Reliability, Responsiveness, Adaptability, Effectiveness, Efficiency, Compatibility,} and \emph{Performance}.

It is worth mentioning that other quality attributes reported in the ISO standards have not been included in CHMM.
This might be due to the impossibility to adapt the attributes to cloud services (e.g., \textit{Freedom From Risk}, \textit{Customer Satisfaction}) or to the impossibility to measure the attribute at runtime, such as for static attributes about the internal structure of the services (e.g., \textit{Maintainability}).  


\subsubsection*{Defining Monitoring Subgoals}
Monitoring subgoals are high level monitoring objectives that represent the decomposition of monitoring goals. 
The ISO standards provide a refinement of quality attributes into quality sub-attributes that help to support the definition of monitoring subgoals. 
We studied the definitions of these sub-attributes offered by the mentioned standards 
to define  subgoals, then we identified subgoals that are measurable in the context of cloud services. 
Considering the \textit{Reliability} of a service as an example, we found that \textit{Continuity}, \textit{Recoverability} and \textit{Availability} are subgoals that indicate to which degree a service provides its outcomes \textit{consistently and stably}. 
With regards to \textit{Effectiveness} and \textit{Efficiency} monitoring goals, the ISO standards do not refine them into quality sub-attributes. Thus, we did it by ourselves, adding extra subgoals that are not mentioned as quality sub-attributes in the ISO.  Table \ref{table:monitoring-goals-definitions} shows the subgoals of the seven monitoring goals coupled with their definitions.

\subsubsection*{Defining Measurable Cloud Properties}
\begin{figure}[]
	\begin{center}
		\includegraphics[width=0.50\textwidth]{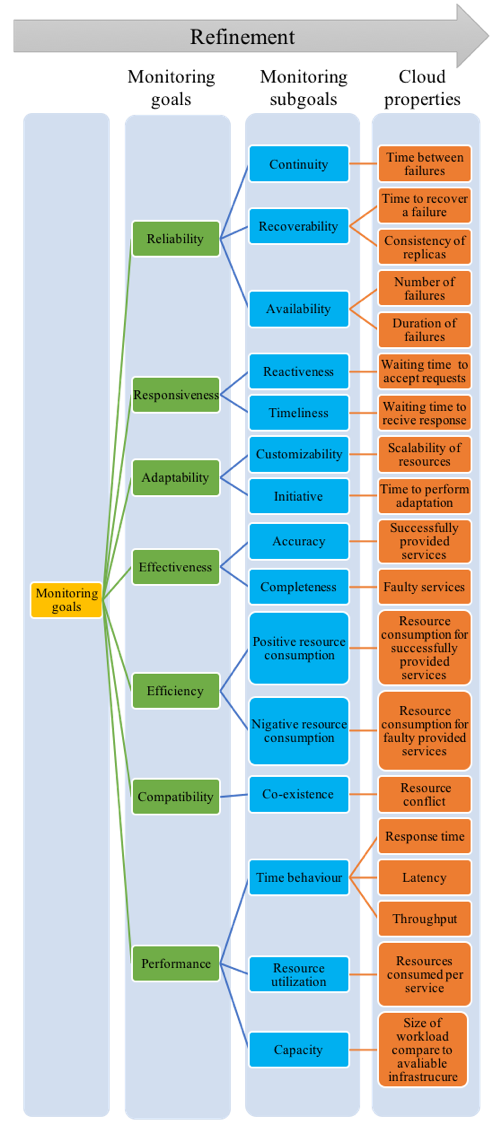}
		\caption{CloudHealth Monitoring Model.}
		\label{fig:quality-model}
        \end{center}
\end{figure}

Cloud properties, such as \textit{memory and CPU consumption}, are measurable characteristics of a cloud system, that is, it is possible to measure them using probes attached to services. 
We mapped each monitoring subgoal to one or more measurable cloud properties. 
We identified cloud properties related to each subgoal based on our experience 
on cloud systems. 
For example, the \textit{Recoverability} of a cloud service is related to two proprieties that are (i) the time required to recover the service in case of failure, and (ii) the consistency of replicas of the service in terms of data and functions. 
We illustrated the resulting CHMM in Figure~\ref{fig:quality-model} in a hierarchical diagram which shows the elements involved in the model and their relationships. 

Since 
CHMM is intended to cover general monitoring goals, there might be the case that some operative scenarios are not covered. 
The model is thus a flexible artifact that can be modified by cloud operators to adapt the CHMM to their needs. 
In other words, an operator can modify or add new monitoring goals and sub-goals and update the mapping of sub-goals to other cloud properties.


\begin{table*}[]
\centering
\begin{tabular}{llp{0.6\textwidth}crl}
\toprule[1pt]
\textbf{Monitoring goal} & \textbf{Subgoal} & \textbf{Definition} \\
\midrule

 \multirow{4}{*}{\textbf{Reliability}} &  \multicolumn{2}{l}{Level of consistency and stability of service outcomes.} \\
  \cmidrule{2-3}
 &  \textbf{Continuity} &  Level of continuity of a service delivered under all expected circumstances, including an acceptable level of mitigation of service interruption risk. \\
   \cmidrule{2-3}

 &  \textbf{Recoverability} &  Level to which a system can restore 
 services (including functions and data) and making them available, in the event of interruptions, failures or disasters.  \\
   \cmidrule{2-3}
   
 &  \textbf{Availability} &  Level of service availability when it is needed. 
 \\
\midrule

 \multirow{ 3}{*}{\textbf{Responsiveness}} & \multicolumn{2}{l}{Level to which the service promptly and timely responds to requests and delivers the required functionalities.}  \\
   \cmidrule{2-3}
 & \textbf{Timeliness} &  Level to which the service timely delivers the required functionalities. \\
   \cmidrule{2-3}

 & \textbf{Reactiveness} &  Level to which the service responds to requests in a prompt way. \\
 \midrule

\multirow{ 3}{*}{\textbf{Adaptability}}   &  \multicolumn{2}{l}{Level of service re-adaptation to meet new needs of users.} \\
     \cmidrule{2-3}
    &  \textbf{Customizability} &  Level of service customization following its user requests.\\
  \cmidrule{2-3}

&  \textbf{Initiative} &  Level at which a service is able to discover users' needs and consequently propose adjustments to meet these needs. \\
 \midrule

\multirow{ 3}{*}{\textbf{Effectiveness}} & \multicolumn{2}{l}{Level of accuracy and completeness that a service achieves in delivering specific goals.} \\
    \cmidrule{2-3}
&  \textbf{Accuracy} &  Level of accuracy that a service achieves in delivering deliver specific goals. \\
  \cmidrule{2-3}

&  \textbf{Completeness} &  Level of completeness that a service achieves in delivering specific goals. \\
  \midrule

\multirow{ 3}{*}{\textbf{Efficiency}} & \multicolumn{2}{l}{Level of resource consumption to deliver specific goals of users in an accurate and complete way.} \\
 \cmidrule{2-3}
&  \textbf{Positive res. consumption} &  Resources consumed to deliver successful services. \\
  \cmidrule{2-3}

&  \textbf{Negative res. consumption} &  Resources consumed for faulty provided services. \\
  \midrule

\multirow{ 2}{*}{\textbf{Compatibility}} & \multicolumn{2}{l}{Level of compatibility of a service with other services while they share the same environment.} \\
\cmidrule{2-3}
&  \textbf{Co-existence} &  Level to which a service is able to efficiently deliver the required functions while it shares hardware or software resources with other services, and without impacting the other services at the same time. \\
  \midrule

\multirow{ 4}{*}{\textbf{Performance}} & \multicolumn{2}{l}{Level of a service performance in relation to the amount of used resources under specific circumstances.} \\
\cmidrule{2-3}

&  \textbf{Time behavior} &  Level to which a service meets the requirements when it delivers its outcomes based on response time, latency and throughput rates. \\
  \cmidrule{2-3}

&  \textbf{Resource utilization} &  Level to which a service meets the requirements when it delivers its outcomes based on the amounts and types of resources. \\
  \cmidrule{2-3}

&  \textbf{Capacity} &  Level to which a service meets the requirements following its maximum limits. \\

\bottomrule[0.7pt]
 \hspace{0.2pt}
\end{tabular}
\caption{Monitoring Goals Model.}
\label{table:monitoring-goals-definitions}
\vspace{-20px}
\end{table*}

\section{Model-Driven Monitoring Process} \label{sec:process}

In this section we describe the CloudHealth process sketched in Figure~\ref{fig:process} more in details.

\subsection{Selecting Monitoring Goals and Services} 
\label{subsec:selection}

Monitoring goals selection is a critical step and requires a certain level of knowledge of both the system and its services. 
Different services might differ in terms of quality demands which corresponds to different KPIs. 
For example, VoIP services are usually characterized by stringent requirements in terms of jitter, but are less demanding 
with respect to bandwidth. On the other hand, video broadcasting services are quite demanding also in terms 
of bandwidth \cite{evans2010deploying}.
%
\ignore{\matteo{lack of coherence between this two paragraphs}}
\ignore{It might be the case that also design decisions might have an impact on the monitoring strategy, since they embrace several aspects, spanning from  security to performance issues. For example, let's say that if we need to provide a mechanism to detect/or even correct on line a message we need to introduce a kind of redundancy. 
A point in case is the use of \textit{parity check} code introduces a control bit that is 0 or 1 depending if the number of bits in the word is even or not. Cyclic redundancy code (CRC) are commonly used in network protocol (e.g., they are present in the TCP datagram, in the Ethernet frame, etc.). 
}
Depending on the services that the operator decides to monitor to watch the health of the system, different sets of KPIs would be collected. To exemplify our approach, we will focus the discussion on the performance aspects (see \textit{Performance} node in CHMM in Figure~\ref{fig:quality-model}) in the next sections.

\subsection{Mapping Monitoring Goals to Software Metrics} \label{subsec:mapping}
\label{subsec:goals-to-metrics}

Once decided the set of interesting monitoring goals for a given operational scenario, 
the second step consists in determining the low level metrics that must be collected to measure the monitoring goals. 
CloudHealth provides the metrics associated to the selected monitoring goals automatically. 

In principle, the process of deriving the metrics from the monitoring goals only requires visiting the tree represented in Figure~\ref{fig:quality-model}, 
from the selected nodes to the leaves. 
As mentioned above, the model is based on ISO standard, which makes it sound. 
At the same time, nothing hampers users from providing their own customization of the model provided that it reflects the hierarchical structure shown in Figure~\ref{fig:quality-model}. 

%
Getting back to our example, following the CHMM, system performance is based on three different dimensions which are: \textit{Time Behaviour}, \textit{Resource Utilization}, and \textit{Capacity}. If we focus on the \textit{Time Behaviour}, we find three metrics that should be collected: \textit{Response time}, \textit{Latency}, and  \textit{Throughput}~\cite{Bilal2013,Persico2015,
	Somasundaram2011,Mann2013,
	Li2017,Zhang2016,
	Naik2017,Moses2011}.

While the three metrics could already capture the time behaviour precisely, operators may decide to change the model to use a specific representation of the time behaviour. For example, the operator may decide to focus on throughput, which is a widely adopted metric that measures the quantity of input being processed by a system in the unit of time. 
In the network management, for example, it can be computed as the average data rate of successful data delivered over a specific communications link\ignore{\matteo{check the definition}}\cite{evans2010deploying}.



\subsection{Identifying the Probes Related to Software Metrics} \label{subsec:identifyingProbes}
In this step CloudHealth automatically identifies the probes that must be used to monitor the identified software metrics on the selected services. For example, in case of performance, CloudHealth will identify a set of probes that can provide information about \textit{Response Time}, \textit{Latency} and \textit{Throughput}. To perform this step, CloudHealth exploits information about the architecture of the cloud system and a catalog of probes that can be used to actually collect software metrics.  

As an example, let us consider a scenario that is becoming more and more popular, where we find a layered 
architecture with a PaaS (Platform as a Service) infrastructure running on top of an IaaS (Infrastructure as a Service) infrastructure, such as AWS~\cite{AWS:18} or Azure~\cite{Azure:18}. 

In this scenario the operators have to decide the granularity that better fits their monitoring strategies. 
Depending on this decision, CloudHealth selects the appropriate probes. For instance, operators may collect CPU and memory utilization at the Virtual Machine level, and thread (e.g., the number of concurrent threads active) and network activity (e.g., average request rate) at the PaaS level. CloudHealth exploits the probes catalog to select and deploy the right probes to collect metrics at the appropriate level.



It is worth noting that the performance metrics (and the related KPIs) are not independent. On the contrary, are statistically (sometimes causally) related. This implies that monitoring goals have to be considered taking care of multiple KPIs. 
We devise a possible CloudHealth implementation where an explicit definition of the metrics is provided,
and the deployment is performed by configuration managers such as Ansible~\cite{Ansible:18}, Chef~\cite{Chef:18} or Puppet~\cite{Puppet:18}. \ignore{\matteo{configuration files? ansible scripts?}} 


\subsection{Attaching Probes to Services}

Once the probes have been selected for deployment, CloudHealth actually deploys them. 
In this context we identified two situations:
\begin{enumerate}
	\item The monitored service and the corresponding probe are deployed at the same time.
	\item The service is already running when the probe is deployed and attached to it.
\end{enumerate}

\subsubsection{Deploying a service together with the corresponding probe} \label{subsub:deploying}
In this situation, the monitored service is not deployed yet. This allows to define a functional block
that includes both the service and the probes. This functional block can be intended as a configuration file that specifies how the service and the probes must be configured and deployed. As an example, it could be implemented as a Cloud Package in Microsoft Azure~\cite{Azure:18}, or as a Template in Google Cloud~\cite{Google:18}.

When dealing with Virtual Machines (VMs), deploying a functional block translates into creating a VM with the service and the related probe pre-installed. The deployment of the VM will result in the service running and the probe attached to it.

When dealing with containers, we may have the service as a container, thus allowing to create a new container, based on it, in which to deploy the probe. Thanks to this structure we could maintain the ability to deploy the service with or without the probe attached without having the storage overhead of keeping two full deployments as it happens with VMs.

\subsubsection{Attaching a probe to a running service}
If the target service is already running a simple solution could be to un-deploy it and then proceed as in Section \ref{subsub:deploying}. This approach however may not be optimal as it requires the service to be stopped.

A more viable solution is to firstly deploy a new instance of the monitored service together with the probe, exploit the load balancer to migrate the load to the new instance, and finally un-deploy the original service as long as it will be serving no requests.

Finally, it is possible to deploy the probe within the guest that is running the service that must be monitored. To inject the probe into the running system, runtime access to the guest is required, that is, a software agent (or an operator) should log into the guest and deploy the probe.
The choice of the deployment strategy depends on the probes that must be deployed and the architecture of the cloud system.

Regardless of the deployment approach, each probe must be configured properly in order to be able to report the monitored data to the monitoring tool (e.g., an ELK installation). As an example, when deploying services and probes with an Ansible-based solution, this is accomplished exploiting \emph{variables}, which values are then referenced in playbooks using the Jinja2~\cite{Jinja2:18} templating system.

\subsection{Reconfiguring the Dashboard}
Operators usually do not deal directly with the choice of the software probes and consequently with the KPIs values produced by them. Instead, they indicate monitoring goals at a high level of abstraction; it is therefore feasible to assume that they might have interest in obtaining health information about the cloud system at the same level of abstraction. However the software probes provide KPIs values at the level of measurable cloud properties. 

The CHMM allows to map the high level goals into cloud properties measured by collecting KPIs values. Since the CHMM model represents the relationships between the high level monitoring goals and the software metrics, CloudHealth can exploit those relationships in both directions. 

CloudHealth exploits this relationship backward in order to aggregate the low level values into high level measures about the monitoring goals. As an example, consider a user that is interested in monitoring Performance. According to the CHMM, the actual software metrics collected by the monitoring framework are \textit{Response Time}, \textit{Latency}, \textit{Throughput}, and \textit{Size of the Workload}. The dashboard will however report a single score for Performance, computed according to the information in the CHMM. 

CloudHealth also exploits forward relationships represented in the tree. For instance, if the operators like to know more about how the performance score is computed, they may expand the view and check the values of all the subgoals and properties that compose it. This feature may become particularly useful when an anomaly is detected in the system, as it allows one to have an overview of the situation first, but also to expand each monitoring goal to discover which cloud property collected from which service is anomalous.

In order to achieve this capability of flexibly browsing the collected information, CloudHealth reconfigures the dashboard based on the selected monitoring goals and the CHMM. In contrast to existing data visualization tools, which are inflexible, CloudHealth exploits a model-driven approach to present all and only the information relevant to the operator.


\section{Use Cases} \label{sec:useCases}
To demonstrate how to get benefits from CloudHealth, we discuss some use cases. Among several possible actors and usage scenarios, we consider two types of actors (managers and technicians) with four usage scenarios. We selected these actors due to the clear difference in their points view about monitoring goals. We will explain each use case in terms of the \textbf{Actor(s)} that will use the case, the \textbf{Wanted Feature} by these actor(s), \textbf{How} CloudHealth supports this usage scenario, and an \textbf{Example} of the use case.

\subsubsection*{Use Case 1: Monitor Services at Business Level} 

\textbf{Actor:} a manager.
\textbf{Wanted Feature:} she wants to monitor cloud services offered by her company at high level of abstraction. This would allow her to take business decisions without delving into technical details.
\textbf{How:} CloudHealth allows her to select monitoring goals at high level of abstractions (e,g., \textit{Performance} of services) using the CloudHealth Monitoring Model. CloudHealth will automatically generate a set of probes attached to the target services to measure KPIs related to the selected monitoring goals. Next, CloudHealth will generate a customized dashboard that allows the actor to monitor KPIs related to her goals and related subgoals at different level of abstractions. 
\textbf{Example:} Daenerys is a manager who works at the business level and does not have knowledge about technical details. Thus, she does not understand low level KPIs, such as the \textit{data rate of successful data delivery over a specific communications link}. She will be happy to abstract away technical details from such KPIs by providing understandable values related to monitoring goals like \textit{Throughput} and \textit{Performance}.

\subsubsection*{Use Case 2: Specified Monitoring for Multi-actors} 

\textbf{Actor:} a manager and a technician.
\textbf{Wanted Feature:} they want to monitor the same service at the same time while they have different interests.
\textbf{How:} CloudHealth offers a customizable model to select monitoring goals. At the technical level, it attaches the set of probes needed to collect the KPIs of interest from the selected services, regardless of the number and type of actors involved. Then, it will provide a customizable dashboard for each actor following the selected goals. 
\textbf{Example:} 
Varys is a salesman who needs to provide real evidences about the health status of the service to customers that can hardly understand low level KPIs. 
Varys can use CloudHealth to abstract these KPIs in a way that is understandable by these customers, for instance abstracting them to \textit{Throughput} and \textit{Performance} properties.
Jon is a software engineer in the same company and he wants to monitor the same service at the same time. Jon is interested in low level monitoring values related to the infrastructure as his goal is to keep the service running with no single point of failure. In this context, CloudHealth provides each one a different dashboard based on their interests.

\subsubsection*{Use Case 3: Add New Monitoring Goals} 

\textbf{Actor:} a manager or a technician.
\textbf{Wanted Feature:} 
they want the flexibility to redefine the hierarchical decomposition of goals into finer grained characteristics.
\textbf{How:} 
CloudHealth is designed to be extendable, which means that the actors are allowed to add new monitoring goals. 
They only need to: (i) extend CHMM in terms of monitoring goals, subgoals (if exists) and cloud properties, and (ii) identify software metrics and their related probes that will be used to collect KPIs about cloud proprieties. The rest of the work will be automatically performed by CloudHealth. 
\textbf{Example:} 
Samwell is a technician who wants to monitor \textit{Security} which is a quality attribute not included in the current version of CloudHealth. 
He needs to extend CHMM by defining  subgoals (e.g., \textit{Confidentiality}, \textit{Integrity}) and maps them to cloud proprieties, software metrics and probes to collect KPIs.

\subsubsection*{Use Case 4: Dynamically Switching Between Monitoring Goals} 

\textbf{Role:} a manager or a technician.
\textbf{Wanted Feature:} they want monitoring flexibility that allows them to dynamically switch between monitoring goals over time based on current interests.
\textbf{How:} CloudHealth provides abilities to its users to switch between monitoring goals once needed. Changing the monitoring goals implies automatically deploying a new set of probes and reconfiguring the dashboard.
\textbf{Example:} Petyr is interested in monitoring \textit{Reliability} of a provided service. Once a VIP customer is being served by this service, Petyr can also extend his monitoring goals and related dashboard to include \textit{Responsiveness} to monitoring the level that the service promptly and timely responds and provides its outcomes for that VIP customer.



\section{Related Work} \label{related}

\label{sec:related.work}
The research described in this paper covers three distinct but related research areas: cloud quality models, cloud monitoring, and health data visualization.

\textit{Cloud Quality Models.} Cloud quality models aim to assess the quality of service and build trust among cloud stakeholders \cite{Lee:QM:2009,Wen:QM:13,Zheng:QM:14}. The problem of establishing a common language between the stakeholder has been tackled by Chen at al. with a model that is characterized by three dimensions (system resource utilization, service performance, and price) described using an ontology language~\cite{chen:2011}. 
Wang et al. studied the relationship between user satisfaction and QoS policies, with the aim to provide a tool to adjust the utility based on the users' demands~\cite{Wang:2016}. 
Zhou et al. proposed a hierarchical model for the evaluation of the cloud service quality. As in our case, starting from the ISO standards, the authors structured the model in six characteristics (usability, security, reliability, tangibility, responsiveness, and empathy) and sub-characteristics~\cite{Zhou:2015}.
Adjoyan et al. defined a quality model of services by analyzing the most commonly known service definitions in the literature. Their model defined seven characteristics related to the structure and the behavior of services. They mapped these characteristics to a set of properties that can be later measured based on software metrics~\cite{adjoyan2014service}.
Other authors, focused their attention on more specific issues. This is the case of Abdeladim et al., who worked on the characterization of elasticity and scalability~\cite{Abdeladim:2014} or Ranaldo et al. who applied a semantic-based QoS model~\cite{Giallonardo:Zimeo:2007} to improve the brokering of resources in grid computing~\cite{Ranaldo2008}. 
However these models do not deal with the actual deployment of probes for collecting the metrics, nor with the aggregation and visualisation of the monitored data.

\smallskip

\textit{Cloud Monitoring.} Monitoring is a necessary task in cloud environments, e.g., to support capacity planning, resource management, SLA compliance and problem solving~\cite{Aceto:MonitoringSurvey:13}. Cloud monitoring solutions range from generic tools that have been extended for monitoring cloud applications, such as Nagios~\cite{Nagios:18} and Ganglia~\cite{Ganglia:18}, to cloud-specific tools such as Amazon CloudWatch~\cite{CloudWatch:18} and Azure Watch~\cite{Azure:18}. However, most of the current monitoring solutions focus on IT metrics rather than operator-defined characteristics and business goals. CloudHealth aims to bridge this gap with a model-driven approach. Recent results improved the support to automatic configuration and deployment of the monitoring infrastructure~\cite{Calero:MonPaaS:15,Monasca:18}. These solutions could be integrated into CloudHealth to implement the automatic reconfiguration and deployment.

\smallskip


\textit{Health Data Visualization.} There are many commercial and open source platforms for \textit{health data visualization}, such as Splunk \cite{Splunk:18} and Grafana \cite{Grafana:18}. These platforms enable the construction of a number of views to present the observed data, and eventually process it. The aim is to support human understanding of the monitoring results and to facilitate decision making as a response to the observed behavior. Most of these solutions allow one to create custom dashboards even without programming knowledge~\cite{Splunk:18,Grafana:18,Kibana:18}. CloudHealth is complementary to these tools since it exploits the capability to build custom views to create an organized set of views about the health of the system. The dashboard is dynamically defined using the CHHM model, which relates the health of the system to the low level KPIs that are collected.

In conclusion, there are not approaches that use models to automatically configure, deploy and operate a complete monitoring solution. The existing work concentrates on some aspects of this process only. CloudHealth represents a first step toward the definition of a complete model-driven process that starts from user needs, includes the automatic deployment of the monitoring probes, and reaches the point of generating  a customized dashboard. 

\section{Conclusions and Future Work}
\label{sec.conclusion}

In this paper, we presented initial results in the design of CloudHealth, a model-driven monitoring solution for watching the health of cloud services. 
CloudHealth exploits a model, the CloudHealth Monitoring Model (CHMM), to identify the software metrics that must be collected based on the high-level monitoring goals selected by the operator. The current version of the CHMM includes 22 monitoring goals organized in a hierarchical structure, where elements at different levels of abstraction are arranged (i.e., main-goals, subgoals, etc.). To derive these monitoring goals, we started from existing standard definitions of quality characteristics that are reported in \textit{IT service quality}, \textit{product quality} and \textit{quality in use} models described in ISO 
standards. 
We further refined these characteristics into measurable cloud properties and concrete software metrics to fully support automatic monitoring of the goals.

Although the current version of CHMM is not intended to be a model that fits all the cases, we design CloudHealth in a flexible way that allows cloud operators to adapt CHMM to their needs.
%
CloudHealth uses CHMM to automatically: (i) control probes that have to be attached to cloud services based on software metrics related to the identified cloud proprieties, (ii) derive goal values from the dynamically collected KPIs, and (iii) automatically generate a dashboard that presents the data coherently with the model. 

We also presented challenges -- and sketch their solutions -- to automatically deploy probes to running services and generate a customizable dashboard that reports KPIs for different cloud operators based on their monitoring goals. We presented four use cases of CloudHealth to show possible usage scenarios at business and technical levels. 
%
 %
 
We plan to extend our work in the following directions.

\begin{description}
\item[Developing a visual environment.] This visual environment will allow cloud operators to work with CloudHealth at different steps, customizing their dashboards and adapting CHMM when needed.

\item[Evaluating CloudHealth] Assessing CloudHealth in real cloud environments, for instance considering both virtual machines and docker containers. 

\item[Extending CloudHealth to different scenarios] Extending\linebreak CloudHealth to support several domains. More precisely, we will consider the context of NGPaaS (Next Generation Platform as a Service, a running H2020 project) where three different cloud scenarios are considered: IoT, Telco and 5G.

\item[Adapting CloudHealth to a Dev-For-Operations model.] Adapting CloudHealth to support Dev-For-Operations model where multi-cloud operators are involved in both the development and operation of a cloud system.
\end{description}

\begin{small}
\section*{Acknowledgment}
This work has been partially supported by the H2020 Learn project, which has been funded under the ERC Consolidator Grant 2014 program (ERC Grant Agreement n. 646867); by the Italian Ministry of Education, University, and Research (MIUR) with the PRIN project GAUSS (grant n.~2015KWREMX), and by the H2020 NGPaaS project (grant n.~761557).
\end{small}

\balance
\bibliographystyle{ACM-Reference-Format}
\bibliography{soheal2018} 

\end{document}